\documentclass[doublecol]{epl2} 

\usepackage{graphics}
\usepackage{subfigure}

\title{Improved community structure detection using a modified fine tuning strategy}
\shorttitle{Improved community structure detection} 

\author{Yudong Sun\inst{1} \and Bogdan Danila\inst{1} \and Kre\v{s}imir Josi\'{c}\inst{2}\and Kevin E.\ Bassler\inst{1,3}}
\shortauthor{Y. Sun \etal}

\institute{                    
  \inst{1} Department of Physics, University of Houston, Houston TX 77204-5005 \\
  \inst{2} Department of Mathematics, University of Houston, Houston TX 77204-3008 \\
  \inst{3} Texas Center for Superconductivity, University of Houston, Houston TX 77204-5002
}


\abstract{
The community structure of a complex network can be determined by
finding the partitioning of its nodes that maximizes modularity.
Many of the proposed algorithms for doing this work by recursively bisecting
the network.
We show that this unduely constrains their results,
leading to a
bias in the size of the communities they find
and limiting their effectivness.
To solve this problem,
we propose adding a step to the existing algorithms
that does not increase the order of their computational
complexity.
We show that, if this step is combined with a commonly used method,
the identified constraint and resulting bias are removed, and its
ability to find the optimal partitioning is improved.
The effectiveness of this combined algorithm is also
demonstrated by using it on real-world example networks.
For a number of these examples, it achieves the best results
of any known algorithm.
}

\begin{document} 
\maketitle 
\noindent 
The problem of detecting and characterizing the community structure of complex networks
has recently attracted a great deal of interest~\cite{Girvan, NewmanSIAM, NewmanPRE69BTW, Radicchi, Variano, Danon, ReichardtPRE, FortunatoPNAS, Weinan, ArenasNJP10}. 
The structure of a network is important for how it functions~\cite{NewmanSIAM}.
Often networks have a modular structure in which the nodes can be partitioned 
into highly connected groups that have
common properties or work together to achieve specific behavior.
This is true for many different types of networks including
social, biological, and telecommunication networks. 
For example, recent work suggests the importance of modularity for the functioning of 
metabolic and transcriptional regulatory networks~\cite{Kreimer, Segal, Snel, Campillos}. 
However, identifying the various communities or modules in a network, 
i.e.\ its community structure, can be difficult. 
This is especially true for large networks which can frequently
be partitioned in multiple, almost equally good ways.
In these cases it becomes very important to eliminate or reduce all biases and 
undue constraints from the method used to detect communities.

Various algorithms have been proposed for the partitioning of complex networks into 
communities~\cite{NewmanPRE69BTW, Radicchi, ArenasNJP10, Boettcher, NewmanPRE69, Donetti, Clauset, FortunatoPRE, ReichardtPRL, Wu, Medus, Massen, Ziv, Duch, Palla, NewmanPNAS, NewmanPRE74, Hastings, Boccaletti, Rosvall-104, Zhang, Lehmann, ArenasNJP9, Sales, Xu, Ruan, Schuetz, Rosvall-105, Agarwal, Nicosia}. 
Many of these algorithms are based on the idea of maximizing a measure called 
modularity $Q$ ~\cite{NewmanPRE69BTW}. 
Finding the partitioning with the absolute maximum modularity has been proven 
to be an NP-hard problem~\cite{Brandes}, and is thus intractable for large networks. 
It is therefore of interest to find a fast algorithm for obtaining a
network partitioning that nearly maximizes modularity.
One of the best methods currently in use is the leading eigenvalue method 
introduced by Newman~\cite{NewmanPNAS, NewmanPRE74}. 
This is a bisectioning method, like most other proposed methods.
It begins by dividing the nodes into two communities and then
recursively divides each of those communities in two
until no further increase modularity can be achieved. 
Each bisection is done by first making 
a ``guess'' at the best way to divide the community into two, 
followed by a ``fine-tuning" that systematically considers moving each of the nodes to 
the other community until no further improvement in modularity can be achieved. 
Note that once a community is generated and fine-tuned the partition between it and 
all other communities  is fixed. It can only be divided further into smaller communities;
none of its nodes are ever placed in other existing communities. 
As we will see, this constraint generates a bias in the results that limits the 
modularity of the resulting partitioning. 

To address this problem, we introduce an additional tuning step in the
algorithm. This additional tuning,  
which is done each time after all communities have been
bisectioned and fine-tuned, systematically considers 
moving each node of the network to all other existing communities or into a new 
community of its own. Using this additional ``final-tuning'', the value of 
modularity is consistently improved. 
The computational cost of final-tuning
is very modest and the algorithm can be used to study large networks.
To demonstrate its usefulness we will use it to
determine the community structure of some known example networks.
As we will see, the results indicate that our algorithm is competitive with 
the best existing community detection methods, and for some networks gives 
partitionings with modularity larger than that obtained
with any other known method.
\begin{figure}
    \scalebox{0.32}[0.32]{\includegraphics*{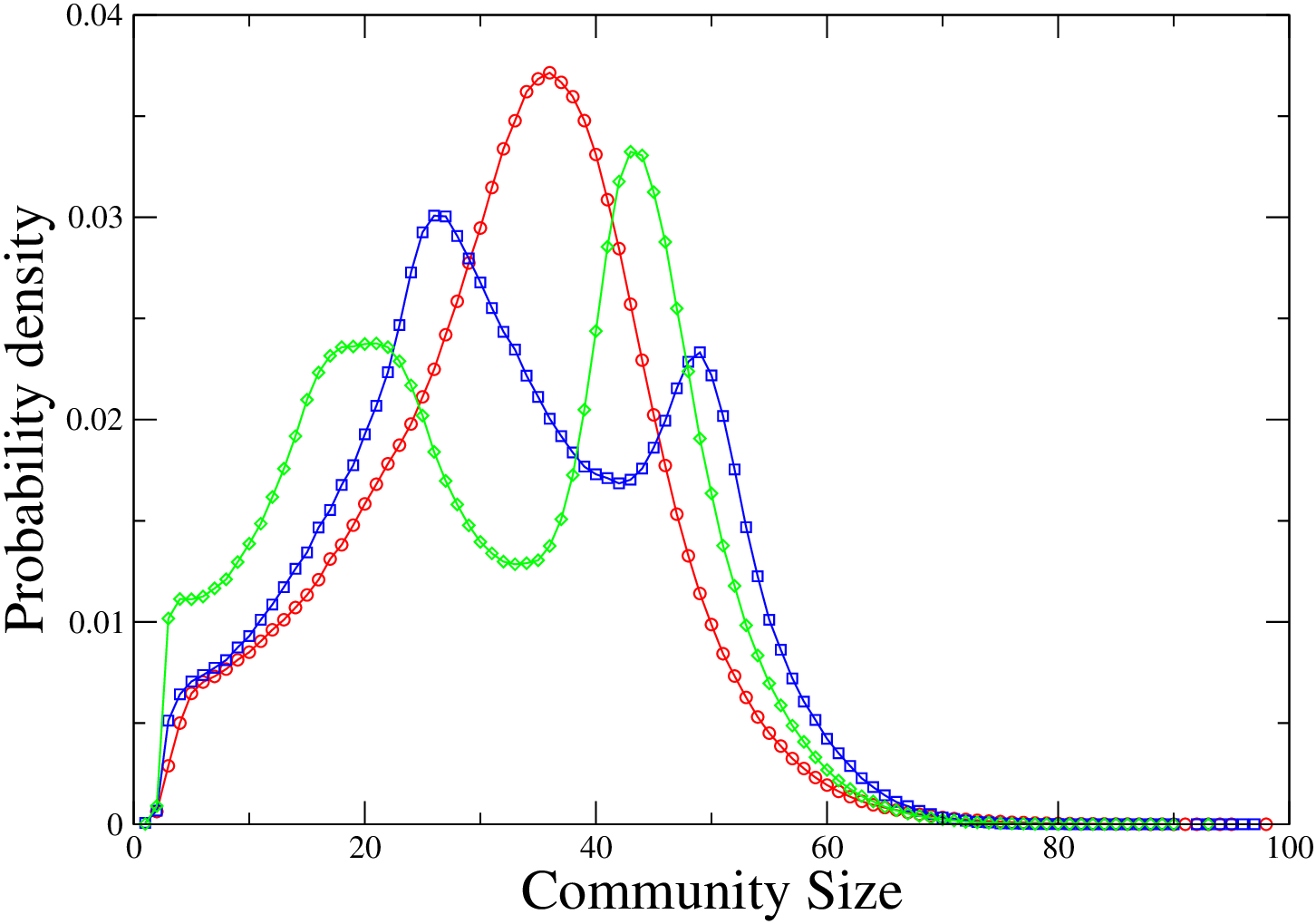}}
    \caption{(Color online) Community size distribution for 
    Erd\H{o}s-R\'enyi random networks with 400 nodes and average degree 4. 
    The results were obtained using the $q=2$ leading 
    eigenvector method with fine-tuning, which is a bisectioning algorithm, (blue squares), 
    the $q=3$ trisectioning variant of that algorithm (green diamonds), 
    and our improved algorithm that combines 
    the $q=2$ variant with ``final-tuning'' (red circles).}
    \label{comm_size}
\end{figure}

Modularity $Q$ is defined as~\cite{NewmanPRE69BTW}
\begin{equation}
\label{def.mod}
    Q=\frac{1}{2m}\sum_{i,j} B_{ij} \delta_{C(i),C(j)}
\end{equation}

\noindent where $B_{ij}=A_{ij}-k_i k_j/(2m)$ are the elements of the 
``modularity matrix'' and $C(i)$ is the community to which node $i$ belongs. 
Here $m$ is the total number of links in the network, 
$k_i$ is the degree of node $i$, $A_{ij}$ are the 
elements of the adjacency matrix, and $\delta$ is the Kronecker delta function.
This definition of modularity is based on the 
idea that links between nodes in the same  
community are dense and that there are only sparse links between 
nodes in different communities. 
It is equal to the fraction of links that connect nodes in the
same community minus what that fraction would be on average
if the communities remained fixed 
but the links were randomly distributed.

The leading eigenvalue method~\cite{NewmanPNAS, NewmanPRE74} produces a guess for 
the initial bisectioning of a network by rewriting the modularity $Q$
in terms of a vector $S$ whose elements $S_i$ take the values $\pm 1$ depending 
on which of the two communities node $i$ belongs to. 
Using the fact that $\delta_{C(i),C(j)}=(S_i S_j+1)/2$, Eq.~\ref{def.mod} becomes
\begin{equation}
\label{Qbis}
    Q=\frac{1}{4m} S^T B S.
\end{equation}

\noindent The problem now becomes to choose the vector $S$ that maximizes $Q$.
If the elements of $S$ were unconstrained, then the best choice would
be to make $S$ equal to the eigenvector corresponding to the largest eigenvalue
of $B$. However, since they are constrained to be $\pm 1$, the method
assigns the sign of $S_i$ to correspond to the sign of the $i$th component
of the eigenvector with the largest eigenvalue. The particular form of 
the modularity matrix $B$ makes the power method a fast and efficient
way to calculate the leading eigenvector~\cite{NewmanPNAS, NewmanPRE74}.

To refine the initial guess, a ``fine-tuning" based on the Kernighan-Lin 
algorithm~\cite{Kernighan-Lin} is used. Once a community has been split in two, 
fine-tuning proceeds by computing for each node $l$ the difference in modularity 
$\delta Q$ associated with moving it from its sub-community to the other,
\begin{equation}
\label{deltaQ}
    \delta Q=-\frac{S_l}{m}\sum_{i\ne l} B_{li} S_i.
\end{equation}

\noindent The move resulting in the highest $\delta Q$ (even if negative) 
is performed and recorded. If multiple moves result in the highest $\delta Q$ 
one of them is picked at random. 
The procedure is then repeated, each time considering only the nodes 
that have not yet been moved and recording the new partitioning, 
until each node has been moved once. 
At this stage, the intermediate partitioning with the highest modularity is retained. 
The entire ``fine-tuning" procedure is repeated at least once, 
until no further increase in modularity can be achieved. 
The move resulting in the largest modularity is then made, 
but, if the modularity after fine-tuning is less than the modularity before, 
the partitioning given by the initial guess is restored.
Subsequent bisectionings of existing communities are done with an analogous
procedure, except that instead of maximizing $Q$ they attempt to maximize
\begin{equation}
\label{DeltaQ}
    \Delta Q=\frac{1}{4m}S^T B^{(C)} S,
\end{equation}

\noindent where
\begin{equation}
\label{BC}
    B^{(C)}_{ij}=B_{ij}-\delta_{ij}\sum_{k \in C}B_{ik}
\end{equation}

\noindent is the modularity matrix for community $C$ and $i,j \in C$.

As far as maximizing modularity is concerned, the leading eigenvalue method 
arguably offers the best balance between speed and accuracy of all methods currently 
in use. 
It is inferior only to simulated annealing but much faster~\cite{Danon}, 
and, for some networks, to greedy algorithms~\cite{Schuetz, Danon} as discussed below. 

However, there is a bias in the method that limits its effectiveness. 
This bias can be demonstrated by considering the distribution of community
sizes that it finds in an ensemble of random networks. 
The blue squares in Fig.~\ref{comm_size} show 
the community size distribution obtained using the leading eigenvalue 
method with fine-tuning for an ensemble of $10^6$ 
random (Erd\H{o}s-R\'enyi) networks. 
The size of these networks is $N=400$ and their average degree is $k=4$. 
To avoid complications with the community detection process, only fully 
connected networks have been considered. The distribution has two 
peaks, one around $N/8$, and another one around $N/16$. Thus, the
method is biased in favor of finding communities of those sizes.
Since, 8 and 16 are both powers of 2, this suggests that
the bias is due to the bisectioning nature of the algorithm.

In order to check if this is true, we introduce a generalized version 
of the leading eigenvector method inspired by the Potts model~\cite{Potts} 
that recursively divides a network 
into $q$ subsets. For $q=2$ it is equivalent to the method described above.
In the generalized method
each of the elements of $S$ are from a set $\{P_1,...,P_q\}$ of $(q-1)$-dimensional 
unit vectors such that $[(q-1)P_i\cdot P_j+1]/q=\delta_{i,j}$. 
These are the vertices of a regular $q$-simplex centered at the origin. 
Modularity is now written as
\begin{equation}
\label{Q_q-sect}
    Q=\frac{q-1}{2mq}\, S^T B S,
\end{equation}

\noindent with the multiplication between elements of $S$ understood 
as a dot product. Equations~\ref{deltaQ} and~\ref{DeltaQ} 
are generalized similarly. The initial guess for $S$ is made 
assuming that an eigenvector of $B$ (or $B^{(C)}$) is equally likely 
to point in any direction in $\mathcal{R}^N$. Hence, normalized 
eigenvectors are uniformly distributed on the $(N-1)$-hypersphere 
of radius one, and the probability density for any component of 
an eigenvector is given by
\begin{equation}
\label{prob.dens}
    p_N(x)=\frac{1}{I_{N-2}}\left(1-x^2\right)^{\frac{N-3}{2}},
\end{equation}

\noindent where $I_N=\sqrt{\pi}\,\Gamma(N/2+1/2)/\Gamma(N/2+1)$. 
The variance of this density is $1/N$. 
In the limit of large $N$, this approaches a Gaussian distribution
whose cumulative probability function will be denoted by $F_N(x)$. 
Now, divide the real axis into $q$ intervals defined 
by $\mathcal{I}_j=\{x|(j-1)/q\le F_N(x) < j/q\}$ and arrange 
the set of vectors $\{P_j\}$ in increasing order of their number 
of positive components. 
Then, if $U$ is the eigenvector corresponding to the largest 
eigenvalue, the initial guess for the $q$-partitioning is
made by choosing $S_i=P_j$ if $U_i\in \mathcal{I}_j$.
Fine-tuning is easily generalized as well. This is done, once a community
has been split into $q$ sub-communities, simply by considering moving each node
from its guessed sub-community to each of the other $q-1$ possibilities.

Here we consider results of only the $q=3$ variant of the generalized
method. This is a trisectioning algorithm. 
In the $q=3$ case, the elements of $S$ are from the set 
$\{(1,0),(-1/2,\sqrt{3}/2),(-1/2,-\sqrt{3}/2)\}$ and 
$\delta_{C(i),C(j)}=(2S_i\cdot S_j+1)/3$. 
If $U$ is the eigenvector corresponding to the largest eigenvalue, 
we choose $S_i=(1,0)$ if $F_N(U_i)\ge 2/3$, $S_i=(-1/2,\sqrt{3}/2)$ 
if $1/3\le F_N(U_i)<2/3$, and $S_i=(-1/2,-\sqrt{3}/2)$ if $F_N(U_i)<1/3$. 
The green diamonds in Fig.~\ref{comm_size} shows the community 
size distribution obtained using this trisectioning method for the 
ensemble of random networks described above. Two peaks are now seen near $N/9$ and $N/27$,
indicating a bias for finding communities of these sizes.
Since, 9 and 27 are powers of 3,
this bias is due to the trisectioning nature of the algorithm.
This result indicates that the peaks found using bisectioning and
trisectioning are artifacts of the methods, which are constrained to 
only allow division of existing communities
and never move nodes to other existing communities.
The same flaw is undoubtedly present in other community detection 
methods that involve simple bisectioning.

\begin{figure}
    \begin{center}
    \scalebox{0.34}[0.34]{\includegraphics{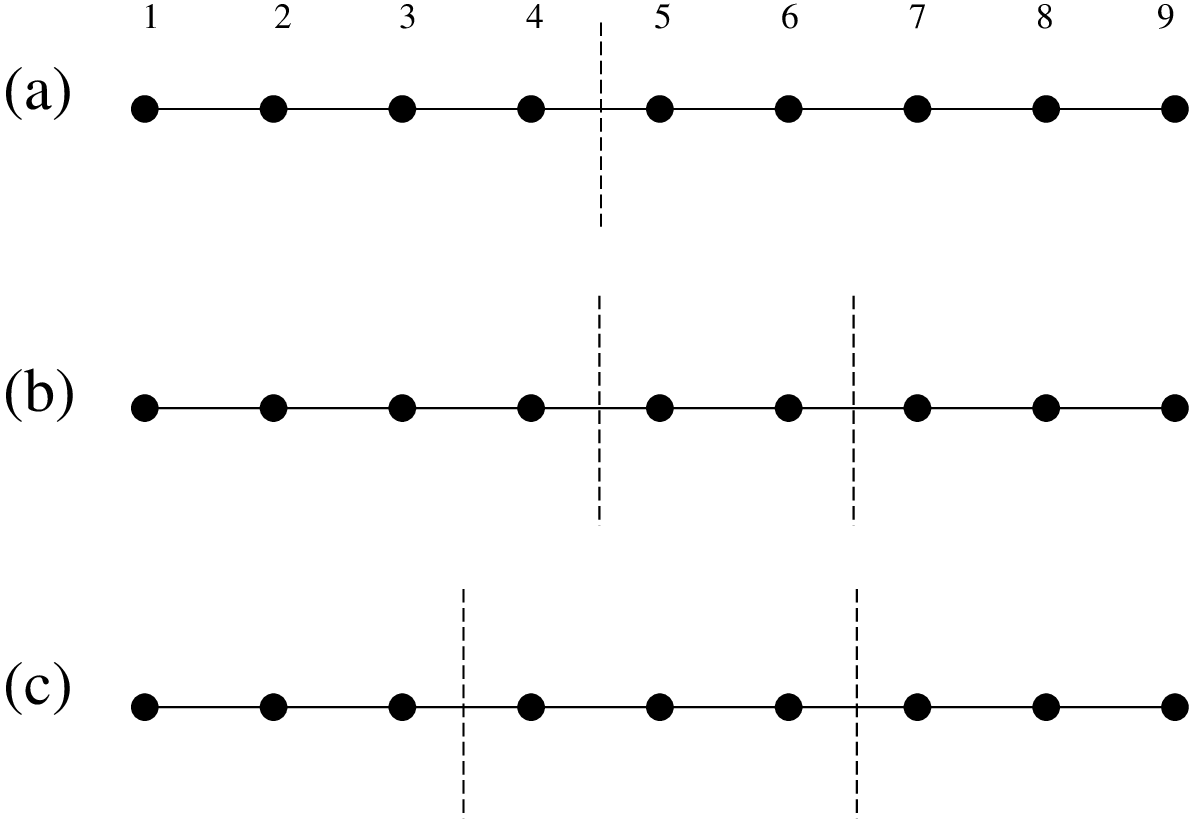}}
    \end{center}
    \caption{Partitions of a simple network (a) after one bisection using any simple bisectioning
    algorithm, (b) final result with simple bisectioning, and (c) the optimal partitioning that
    maximizes modularity. The optimal partitioning results if 
    bisectioning is combined with final-tuning. }
    \label{little_net}
\end{figure}

To better understand the problem, consider the simple example in 
Fig.~\ref{little_net} of a linear network with nine nodes connected by 
eight links. 
Fig.~\ref{little_net}a shows the partitioning after one bisection.
Further bisectioning divides the community with five nodes
into two sub-communities as shown in Fig.~\ref{little_net}b. 
The modularity of this configuration is $Q=51/128$. 
However, this is not the partitioning that maximizes modularity. 
The partitioning shown in Fig.~\ref{little_net}c 
exhibits the largest modularity $Q_{max}=53/128$. 
This partitioning will never be found with a simple bisectioning algorithm.
This is because nodes 4 and 5 
are assigned to different communities at the first bisectioning, and
can never be moved into the same community after that. 
However, they are in the same community in the optimal partitioning.
In more general terms, if the partitions set during a division  
are not optimal, there is no chance to correct them. 	
These partitions unduly constrain the community detection algorithm.

The problem can be solved by using an additional modified fine-tuning step 
similar to the one described above but involving all partitions. 
This modified fine-tuning, which we call ``final-tuning,'' 
is performed at the end of every round of divisions in the course of which 
all communities that resulted from the previous round have been divided once. 
In the final-tuning procedure, the 
differences in modularity associated with moving 
every node of the network from its sub-community to every other sub-community are computed, 
and the move resulting in the highest modularity difference is performed and recorded. 
As in the case of fine-tuning, if multiple moves result in the highest 
modularity difference one of them is picked at random. 
The procedure is repeated on the nodes that have not yet been moved 
until each node has been moved once, at the end retaining the intermediate configuration 
with the highest modularity. 
Equation~\ref{deltaQ} can be used to calculate the difference in modularity 
$\delta Q$ when switching one node to an arbitrary community, by 
defining community $C$ as the union of the origin and destination communities. 
However, it is usually faster to calculate $\delta Q$ directly from 
Eq.~\ref{def.mod}~\cite{ReichardtPRE}. That is,
by subtracting the modularity calculated with node $l$ in 
community $Y$ from the modularity calculated 
with $l$ in its original community $X$ we find
\begin{equation}
\label{deltaQfin}
    \delta Q=\frac{1}{m}\left(\sum_{i\in Y} A_{li}-\sum_{i\in X} A_{li}-\frac{k_l (K_Y-K_X+k_l)}{2m}\right),
\end{equation}

\noindent where $K_X$ and $K_Y$ are the sums of the degrees of all nodes in 
communities $X$ and $Y$ respectively, and $k_j$ is the degree of node $j$.
The speed of final-tuning can be improved by only considering moving nodes to
the communities of the nodes which they are connected to, or into a community of
their own.

To incorporate final-tuning into the community detection, 
we suggest the following algorithm.

\begin{enumerate}
\item Apply any previous method, for example the leading eigenvalue method with fine-tuning,
to attempt to divide each of the existing communities.

\item Use Eq.~\ref{deltaQfin} to calculate $\delta Q$ caused by all 
possible moves of any single node to all other existing communities,
or into a community of its own.

\item Find the move that leads to the largest $\delta Q$ (even if negative) and make the move. 
If multiple moves result in the largest $\delta Q$ pick one of them randomly. 
Fix the community assignment for the node moved.

\item Repeat steps 2-3, but in step 2 only consider moving nodes whose
community assignment has not yet been fixed. Continue
repeating until every node is moved once and only once.

\item Choose the intermediate configuration with the largest $Q$.

\item Repeat 2-5 until no further improvement of $Q$ is achieved.

\item Compare the modularity of the resulting partition to the modularity after step 1. 
If the latter is larger, then revert to the partitioning after step 1.

\item Repeat 1-7 until no further improvement of $Q$ is achieved.
\end{enumerate}

The computational complexity of the leading eigenvalue algorithm with fine-tuning is 
$O[N(N+m)]$ for one bisection~\cite{NewmanPNAS}. 
Since the expected number of bisections is $O(\log N)$, the overall complexity of the 
community detection procedure is $O[N(N+m)]\log N$~\cite{NewmanPNAS}. 
The additional workload due to investigating all possible moves between communities 
in a final-tuning step is generally $O[N^2]$. 
Thus, if the leading eigenvalue algorithm with fine-tuning is used in step 1, 
then combining it with final-tuning does not change the overall order 
of computational complexity.
\begin{figure}
    \scalebox{0.36}[0.36]{\includegraphics*{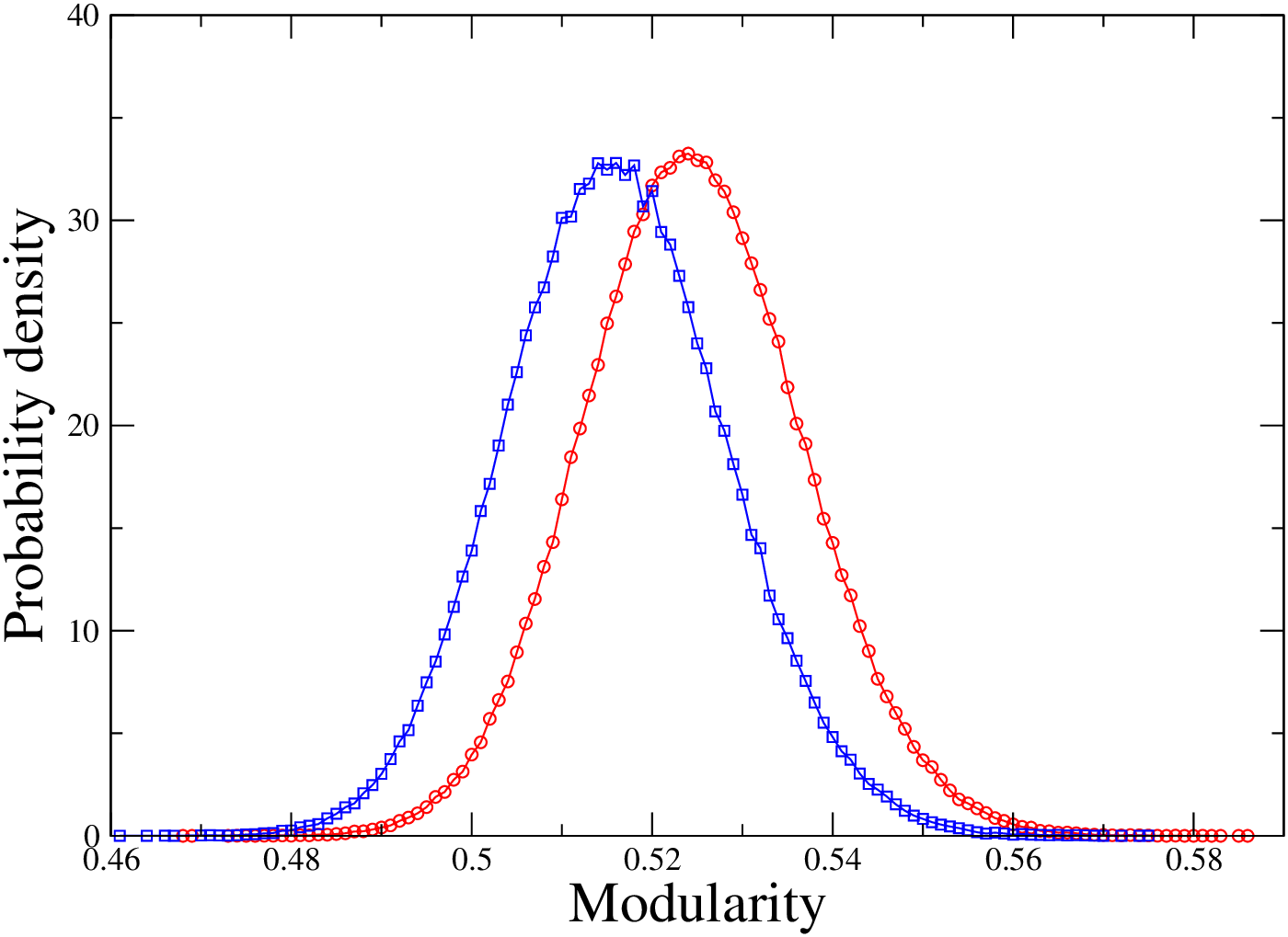}}
    \caption{(Color online) Modularity distribution for 
    Erd\H{o}s-R\'enyi random networks with 400 nodes and average degree 4. 
    The results were obtained using the $q=2$ leading 
    eigenvector method with fine-tuning (blue squares), 
    and by combining it with ``final-tuning'' (red circles).}
    \label{qdist}
\end{figure}

The effectiveness of including final-tuning in a community detection algorithm
can be demonstrated first by noting that if it is used, then the optimal partitioning 
of the simple network in Fig.~\ref{little_net} (shown in Fig.~\ref{little_net}c)
is obtained. This is because final-tuning removes the undue constraints 
that exist in simple bisectioning methods. 
Second, combining final-tuning with the $q=2$ leading
eigenvalue method and fine-tuning, 
the multiple peaks of the community size distribution 
for random networks disappear (Red circles in Fig.~\ref{comm_size}), leaving
only a single peak as expected~\cite{ReichardtPRE}. 
This indicates that the undue constraints have been removed.
The resulting modularity is also significantly improved. 
Figure~\ref{qdist} compares the distribution of modularity of partitionings 
obtained with (red circles) and without (blue squares) final-tuning 
for the same ensemble of random networks. 
In both cases, the results approximately follow a Gaussian distribution, 
but the average maximum modularity increases 
from $0.515748\pm 0.000067$ to $0.524293\pm 0.000057$ with the use of the 
finial-tuning. The standard deviations of the two distributions 
are $0.012158\pm 0.000082$ and $0.012022\pm 0.000068$, respectively. 
All errors are $2\sigma$ estimates.
Note that using the leading eigenvalue method with $q>2$ does not
improve the $q=2$ results. (Data not shown.)
\begin{table}
\caption{Comparison of maximum modularity of some example networks
found using our method of final-tuning combined with
the $q=2$ leading eigenvalue method and fine-tuning, $Q_{FT}$,
against the best previously known result obtained using any other algorithm, $Q_{pub}$.
The source of the data for the network is cited next to its name, and the articles
reporting the previous best results are cited under ``Method.''}
\label{tab.1}
\begin{center}
\begin{tabular}{ccccc}
\hline
\hline
Network & Size & $Q_{FT}$ & $Q_{pub}$ & Method \\
\hline
Karate\cite{Zachary} & 34 & 0.420 & 0.420 & \cite{Agarwal} \\
Jazz musicians\cite{Gleiser} & 198 & 0.445 & 0.445 & \cite{Duch, Agarwal} \\
Metabolic\cite{Jeong} & 453  & 0.452 & 0.450 & \cite{Schuetz, Agarwal} \\
E-mail\cite{Guimera} & 1133 & 0.580 & 0.579 & \cite{Agarwal} \\
Key signing\cite{Guardiola,Boguna} & 10680 & 0.867 & 0.878 & \cite{Schuetz} \\
Physicists\cite{NewmanPNAS2001} & 27519 & 0.737 & 0.748 & \cite{Schuetz} \\
\hline
\hline
\end{tabular}
\end{center}
The networks are, respectively, the karate club of Zachary, a collaboration
network of early Jazz musicians, a metabolic network of the nematode {\it Caenorhabditis 
elegans}, a network of e-mail contacts at a university, a trust network of mutual 
cryptography key signings, and a coauthourship network of condensed matter physicists.
These same networks were studied in Ref.~\cite{NewmanPNAS}.
\end{table}

As a further test, we applied final-tuning, 
combined with the $q=2$ leading eigenvalue method and fine-tuning, 
to find the community structure of a number of real world example
networks that have been studied in the literature~\cite{Schuetz, Duch, NewmanPNAS, 
Agarwal, Weinan, Zachary, Gleiser, Jeong, Guimera, Guardiola, Boguna, NewmanPNAS2001}. 
Note that, because the algorithm is partly stochastic, we have run hundreds of analyzes
for each of the networks considered and chosen the partition with the largest modularity. 
Table~\ref{tab.1} shows a comparison between our results (labeled ``FT") and the 
best known published results (labeled ``pub"). 
Although the increases in modularity may only seem modest, the estimated
upper bounds for the maximum modularity for some of these networks~\cite{Agarwal} 
are very close to our results. 
For the two smallest networks, we find what is likely to be the optimal
partitioning. For the two medium sized networks, our algorithm achieves
results better than any other known algorithm. However, for the two largest 
networks a greedy algorithm performs better. Greedy algorithms work
by combining communities, instead of dividing them. It is possible that
combining a greedy algorithm with final-tuning would improve the results
for the two largest networks as well. Note that for all examples studied
our results using final-tuning combined with the leading eigenvalue method
equaled or surpassed the best known results obtained using the leading
eigenvalue method only.

In summary, we have shown that an undue constraint exists in many of the algorithms
commonly used to detect the community structure of 
complex networks, and that this constraint limits the effectiveness of those algorithms
for finding the network partitioning with the maximum modularity. 
To solve the problem, we proposed adding an extra step into the community detection
procedure that fine-tunes the partitioning by
considering moving nodes to all other existing communities.
This extra step requires only minor computational cost, and does not
increase the order of the computational complexity of the overall algorithm.
We demonstrated the effectiveness of our improved algorithm by 
showing that it eliminates the bias towards dividing random 
networks into communities of size $N/2^n$
that exists when a bisectioning algorithm is used.
We also used it to determine the community structure of some known example networks
and found that, except for the largest networks, it achieves the best
results of any currently known algorithm. Finally, since finding the network
partitioning with the largest modularity is an an NP-hard problem~\cite{Brandes},
we note that the approach we have used here to improve the approximate algorithms
for solving the problem, namely of identifying and removing undue constraints that
bias their results, may be helpful in improving approximate algorithms for solving 
other NP-hard problems associated with complex networks 
such as finding the set of routes that maximizes the capacity of congested network
transport~\cite{OurPRE, OurChaos, OurEPL}.

\acknowledgments
The work of YS, BD, and KEB was supported by the NSF through grant No.\ DMR-0427538,
and by the Texas Advanced Research Program through grant No.\ 95921.
The work of KJ was supported by the NSF through grants No.\ DMS-0604429 and
No.\ DMS-0817649, and by the Texas Advanced Research Program through grant
No.\ 96105.
The authors gratefully acknowledge Tim F.\ Cooper and Charo I.\ Del Genio
for stimulating discussions.

\end{document}